\documentclass[aps]{revtex4}
\usepackage{amssymb}
\usepackage{amsmath}
\usepackage{bm}
\usepackage{graphicx}
\usepackage{epsfig,amsmath}

\begin{document}

\title{Langevin dynamics of vortex lines in the counterflowing He II. Talk given at the Low Temperature Conference, Kazan, 2015  }
\author{Sergey K. Nemirovskii\thanks{%
email address: nemir@itp.nsc.ru}}
\affiliation{Institute of Thermophysics, Lavrentyev ave, 1, 630090, Novosibirsk, Russia
and Novosibirsk State University, Novosibirsk}
\date{\today }

\begin{abstract}
The problem of the statistics of a set of chaotic vortex lines in a
counterflowing superfluid helium is studied. We introduced a Langevin-type
force into the equation of motion of the vortex line in presence of relative
velocity $\mathbf{v_{ns}}$. This random force is supposed to be Gaussian
satisfying the fluctuation-dissipation theorem. The corresponding
Fokker-Planck equation for probability functional in the vortex loop
configuration space is shown to have a solution in the form of Gibbs
distribution with the substitution $E\{\mathbf{s\}\rightarrow }E(\{\mathbf{%
s\}-P(v_{n}-v_{s})}$, where $E\{\mathbf{s\}}$ is the energy of the vortex
configuration $\{\mathbf{s\}}$, and $\mathbf{P}$ is the Lamb impulse. Some
physical consequences of this fact are discussed.\\
\newline
PACS numbers: 47.32.C- (Vortex dynamics) 47.32.cf (Vortex reconnection and
rings), 47.37.+q (Hydrodynamic aspects of superfluidity)

\end{abstract}

\maketitle

\section{Introduction and scientific background}

\label{intro}

The Langevin approach and, relating to it, the Fokker-Planck equation are
the very powerful methods for both analytical and numerical studies of
stochastic processes in various fields of theoretical physics \cite%
{Zinn-Justin2002}. These methods are also applied for study of
nonequilibrium states, in particular, for classical turbulence (see for
details the book \cite{McComb1990}). They allow e.g. to develop
renormalization procedure and, modifying the shape of correlator of the
stirring force, to test various physical situations. For instance, if the
stirring force has a correlator proportional to the squared wave vector,
then the pumping locally equilibrates viscous damping and the system is
driven into a thermal equilibrium state. However, when stirring force are
concentrated on large scales (small wave numbers) the according procedure
results in the solution with Kolmogorov cascade (although these results are
not very rigorous and there are many open questions left).

In the dynamics of quantum vortices the Langevin and the Fokker-Planck
approaches were applied first to describe the growth of small ideally
circular vortex rings under thermal activation (see for details and
bibliography the book \cite{Donnelly1991}). For an ensemble of vortices this
theory had been used for dynamic processes in films of $^{4}$He, variant of
nonstationary Kosterlitz-Thouless phase transition \cite{Ambegaokar1980}. A
combination of the above two examples was the use of the Fokker-Planck
equations for gas of circular vortex rings to describe lambda transition in
superfluid helium (for review and bibliography see e.g. \cite{Williams2003}%
). In papers \cite{Nemirovskii1993a},\cite{Nemirovskii1995} there was an
attempt to study both numerically and analytically the stochastic behavior
of single, non-circular vortex loop (disregarding reconnections) in the
local induction approximation and driven by low-frequency noise.

The problem of the equilibrium statistics of a set of chaotic vortex lines
in a superfluid helium had been studied \cite{Flandoli2001},\cite%
{Nemirovskii2004},\cite{Nemirovskii2009} some time ago. Considering vortex
filament undergoing random forcing, the Langevin-type equation of motion of
the line had been obtained. The respective Fokker-Planck equation for
probability functional $\mathcal{P}(\{\mathbf{s}(\xi )\})$ in vortex loop
configuration space is shown to have a solution of the form $\mathcal{P}(\{%
\mathbf{s}(\xi )\})=\mathcal{N}\exp (-H\left\{ \mathbf{s}\right\} /T),$
where $\mathcal{N}$ is a normalizing factor and $H\left\{ \mathbf{s}\right\}
$ is energy of vortex line configurations.

In the present paper we study dynamics of the vortex filament under action
of the Langevin-type random force in presence of relative velocity $\mathbf{v%
}_{ns}$ . Motivation of this topic is due to fact that chaotic set of vortex
filaments, the so called quantum turbulence develops in the counterflowing
superfluid helium without random stirring, so it is important to compare
both mechanisms.

\section{Fokker-Planck equation}

\label{FP}

We consider the Langevin dynamics of vortex loops in three-dimensional space
with no boundaries. The equation of motion of the vortex line elements reads
\begin{equation}
\mathbf{\dot{s}=B(}\xi )+\mathbf{v}_{s}+\alpha \mathbf{s}^{\prime }(\xi
)\times (\mathbf{v}_{ns}-\mathbf{B(}\xi ))+\alpha ^{\prime }\mathbf{s}%
^{\prime }(\xi )\times \mathbf{s}^{\prime }(\xi )\times (\mathbf{v}_{ns}-%
\mathbf{B(}\xi ))+\mathbf{f(}\xi _{0},t).  \label{ds/dt}
\end{equation}

Here $\mathbf{B(}\xi )$ is the self induced, Biot-Savart, propagation
velocity of the vortex filament at a point $\mathbf{s}$, determined by
Biot-Savart law; $\mathbf{s(\xi ,t)}$ is the radius-vector of the vortex
line points; $\xi $ is a label parameter, in this case it coincides with the
arc length; $\mathbf{v}_{ns}$ is the relative velocity between the normal
and superfluid component helium; $\mathbf{s^{\prime }}$ is the derivative
with respect to the arc length, $\alpha $, $\alpha ^{\prime }$ are the
friction coefficients, describing interaction of vortex filament with normal
component, and $\mathbf{f(}\xi ,t\mathbf{)}$ is the Langevin force. The
Langevin force is supposed to be a white noise with the following correlator
\begin{equation}
F=\left\langle \mathbf{f}_{i}(\xi _{1},t_{1})\mathbf{f}_{j}(\xi
_{2},t_{2})\right\rangle \ =\frac{2k_{B}T\alpha }{\rho _{s}\kappa }\delta
_{ij}\delta \ (t_{1}-t_{2})\delta \ (\xi _{1}-\xi _{2}).  \label{correlator}
\end{equation}%
Here $i,j$~are the spatial components; $t_{1},t_{2}$ are the arbitrary time
moments; $\xi _{1},\xi _{2}$ denote any points on the vortex line. The
\textquotedblleft temperature\textquotedblright , related to this force via
the fluctuation dissipation theorem is an artificial quantity, having
nothing to do with real temperature. In this sense we study a model problem,
which has application to a number of problems on chaotic vortex lines,
including superfluid turbulence. We can claim, that it would be considered
as 3D variant of the famous Onsager model \cite{Onsager1949}. The only case,
where this approach can be applied for real situation, (at least for quantum
fluids) is the vicinity of phase transition (due to the fact, that the
superfluid density is very small there).

There is one specific feature. The term with $\alpha ^{\prime }$\ doesn't
enter into the fluctuation dissipation theorem. Its role is reduced to that,
that it just changes the self-induced velocity (Biot-Savart term), but
doesn't affect the effective temperature. It could be said that for
dissipation of energy only the $\alpha $\ term is responsible, while the
reactive mutual-friction $\alpha ^{\prime }$\ just slightly renormalizes the
inertial term of conventional hydrodynamics. This situation (for different
reason) had been discussed earlier in paper \cite{Finne2003}. Further we
omit term with $\alpha ^{\prime }$\ .

Let us consider the probability distribution functional \cite%
{Nemirovskii2004} defined as
\begin{equation}
\mathcal{P}(\{\mathbf{s}(\xi )\},t)=\left\langle \delta \left( \mathbf{s}%
(\xi )-\mathbf{s}(\xi ,t)\right) \right\rangle .  \label{pdf}
\end{equation}%
The Fokker-Planck equation for the time evolution of quantity $\mathcal{P}(\{%
\mathbf{s}(\xi )\},t)$ can be derived from equation of motion (\ref{ds/dt})
in standard way (see e.g. \cite{Zinn-Justin2002},\cite{Nemirovskii2004} )%
\begin{equation}
\frac{\partial \mathcal{P}}{\partial t}+\int d\xi \frac{\delta }{\delta
\mathbf{s}(\xi )}{\Huge \{}\left[ \mathbf{B(}\xi )+\mathbf{v}_{s}+\alpha
\mathbf{s}^{\prime }(\xi )\times (\mathbf{v}_{n}-\mathbf{v}_{s}-\mathbf{B(}%
\xi ))\right] +F\frac{\delta }{\delta \mathbf{s}(\xi )}{\Huge \}}\mathcal{P}%
\ =0  \label{FP Vns}
\end{equation}

\section{Equilibrium solution to Fokker-Planck equation}

\label{solution}

We are looking for solution of the following form the Gibbs distribution
\begin{equation}
\mathcal{P}(\{\mathbf{s}(\xi )\},t)=\mathcal{N}\exp (-\frac{H\{\mathbf{s\}}}{%
k_{B}T}),  \label{Gibbs 1}
\end{equation}

Where \ the Hamiltonian $H\{\mathbf{s\}}$ is%
\begin{equation}
H\{\mathbf{s\}=}E\{\mathbf{s\}-P(v_{n}-v_{s}).}  \label{hamiltonian}
\end{equation}

The energy $E(\mathbf{s})$ of vortex configuration $\{\mathbf{s\}}$ is

\begin{equation}
E(\mathbf{s})=\frac{{\rho }_{s}{\kappa }^{2}}{8\pi }\int\limits_{\Gamma
}\int\limits_{\Gamma ^{^{\prime }}}\frac{\mathbf{s}^{\prime }(\xi )\mathbf{s}%
^{\prime }(\xi ^{\prime })}{|\mathbf{s}(\xi )-\mathbf{s}(\xi ^{\prime })|}%
d\xi d\xi ^{\prime }.  \label{H(s)}
\end{equation}%
\ \ \ \ On the role of momentum we take the so called Lamb Impulse defined as

\begin{equation}
\mathbf{P}\ \mathbf{=}\frac{\rho _{s}\kappa }{2}\int \mathbf{s}(\xi )\times
\mathbf{s}^{\prime }(\xi )\ d{\xi .}  \label{Lamb(r)}
\end{equation}

In usual statistical mechanics when $\mathbf{P}$ is true momentum of
particle (or quasiparticle) Eq. (\ref{Gibbs 1}) is obvious and follows from
the Galilean transformation. Since the Lamb impulse is not "real" momentum,
then Eq. (\ref{Gibbs 1}) is not obvious and needs in the verification. Let's
give the proof the realization of the Gibbs distribution in form Eq. (\ref%
{Gibbs 1}).

First of all we give two identities. The variational derivative of energy of
vortices is related to the self induced Biot-Savart propagation velocity $%
\mathbf{B(}\xi )$ in following manner

\begin{equation}
\mathbf{B(}\xi )=\frac{1}{{\rho }_{s}{\kappa }}\mathbf{s}^{\prime }(\xi
)\times \frac{\delta E(\mathbf{s)}}{\delta \mathbf{s(}\xi _{0},t)}%
\Rightarrow \rho _{s}\kappa \mathbf{s}^{\prime }(\xi )\times \mathbf{B(}\xi
)=\frac{\delta E(\mathbf{s)}}{\delta \mathbf{s(}\xi ,t)}.  \label{B_trans}
\end{equation}%
The latter equality requires $\mathbf{s}^{\prime }(\xi )\mathbf{B}(\xi )=0,$%
which can be reached by suitable reparametrization. We take parametrization $%
\mathbf{s}(\xi ,t)$, where $\mathbf{s}^{\prime }(\xi )\delta E(\mathbf{s}%
)/\delta \mathbf{s}(\xi ,t)=0$ and $\mathbf{s}^{\prime }(\xi )\mathbf{s}%
^{\prime }(\xi )=1$

Another identity concerns of variational derivative of quantity $\mathbf{%
P(v_{n}-v_{s}).}$For constant vector $\mathbf{v}_{ns}$ the following
relation takes place:
\begin{equation}
\rho _{s}\kappa \mathbf{s}^{\prime }(\xi )\times \mathbf{v}_{ns}=\frac{%
\delta (\mathbf{Pv}_{ns}\mathbf{)}}{\delta \mathbf{s(}\xi ,t)}.
\label{dpv/ds}
\end{equation}%
Now we have to substitute the Gibbs distribution (\ref{Gibbs 1}) in the
Fokker-Planck equation (\ref{FP Vns}). We take that $\mathbf{v}_{n}$ is the
time independent and we work in a frame where $\mathbf{v}_{n}$ is zero. Let
us check the non dissipative terms, the first two terms in square brackets
of the Fokker-Planck equation (\ref{FP Vns}). It is readily verified that

\begin{eqnarray}
&&\int d\xi \lbrack \frac{\delta }{\delta \mathbf{s}(\xi )}(\mathbf{s}%
^{\prime }(\xi )\times \frac{\delta E(\mathbf{s)}}{\delta \mathbf{s(}\xi ,t)}%
+\mathbf{v}_{s})]\exp (-\frac{H(\mathbf{s)}}{k_{B}T})  \nonumber
\label{Nondiss} \\
&&-\frac{1}{k_{B}T}\int d\xi (\mathbf{s}^{\prime }(\xi )\times \frac{\delta
E(\mathbf{s)}}{\delta \mathbf{s(}\xi ,t)}+\mathbf{v_{s}}){\Large (-}\frac{%
\delta E(\mathbf{s)}}{\delta \mathbf{s(}\xi ,t)}+\mathbf{s}^{\prime }(\xi
)\times \mathbf{v}_{s}{\Large )}\exp (-\frac{H(\mathbf{s)}}{k_{B}T})
\nonumber \\
&&
\end{eqnarray}%
Let us check the first line, which includes differentiation of terms inside
of the square brackets%
\begin{equation}
\int d\xi \left\{ \frac{\delta }{\delta \mathbf{s}(\xi )}\left[ \mathbf{s}%
^{\prime }(\xi )\times \frac{\delta E(\mathbf{s)}}{\delta \mathbf{s(}\xi ,t)}%
+\mathbf{v_{s}}\right] \right\} \exp (-\frac{H\{\mathbf{s\}}}{k_{B}T})
\label{nondiss}
\end{equation}

Performing functional differentiation and using tensor notation, the first
term in square brackets is proportional to

\begin{equation}
\epsilon ^{\alpha \beta \gamma }\frac{\delta }{\delta \mathbf{s}_{\alpha
}(\xi )}\left( \mathbf{s}_{\beta }^{\prime }(\xi )\frac{\delta H(\mathbf{s)}%
}{\delta \mathbf{s}_{\gamma }\mathbf{(}\xi ,t)}\right) .
\end{equation}%
The functional derivative $\delta \mathbf{s}_{\beta }^{\prime }(\xi )/\delta
\mathbf{s}_{\alpha }(\xi )\propto \delta _{\beta \alpha }$, therefore this
term vanishes due to symmetry. Further, the functional derivative from
constant velocity obviously also vanishes $\mathbf{v}_{s}$, $\delta \mathbf{v%
}_{s}/\delta \mathbf{s}(\xi )=0$. The rest terms are
\begin{eqnarray}
-\frac{1}{k_{B}T}\int d\xi (\frac{\delta E(\mathbf{s)}}{\delta \mathbf{s(}%
\xi ,t)}(\mathbf{s}^{\prime }(\xi )\times \frac{\delta E(\mathbf{s)}}{\delta
\mathbf{s(}\xi ,t)}\mathbf{)}-\rho _{s}\kappa (\mathbf{s}^{\prime }(\xi
)\times \mathbf{v}_{s})(\mathbf{s}^{\prime }(\xi )\times \frac{\delta E(%
\mathbf{s)}}{\delta \mathbf{s(}\xi ,t)})  \nonumber \\
+\frac{\delta E(\mathbf{s)}}{\delta \mathbf{s(}\xi ,t)}\mathbf{v}_{s}-\rho
_{s}\kappa (\mathbf{s}^{\prime }(\xi )\times \mathbf{v_{s})v}_{s}\mathbf{)}%
\exp (-\frac{H(\mathbf{s)}}{k_{B}T})  \nonumber \\
\end{eqnarray}

Again the first terms vanishes due to symmetry and the forth term is zero,
since production $\mathbf{s}^{\prime }(\xi )\times \mathbf{v_{s}}$ is normal
to $\mathbf{v_{s}}$. The rest ones lead to the following expression (we omit
factor $-1/k_{B}T$, and choose parametrization where and $\mathbf{s}^{\prime
}(\xi )\mathbf{s}^{\prime }(\xi )=1$)%
\begin{eqnarray}
\int d\xi \lbrack \frac{\delta E(\mathbf{s)}}{\delta \mathbf{s(}\xi ,t)}%
\mathbf{v}_{s}-\rho _{s}\kappa (\mathbf{s}^{\prime }(\xi )\times \mathbf{v}%
_{s})(\mathbf{s}^{\prime }(\xi )\times \frac{\delta E(\mathbf{s)}}{\delta
\mathbf{s(}\xi ,t)})]\exp (-\frac{H(\mathbf{s)}}{k_{B}T})=  \nonumber \\
\int d\xi \lbrack -\mathbf{v}_{s}\frac{\delta E(\mathbf{s)}}{\delta \mathbf{%
s(}\xi ,t)})+(\mathbf{s}^{\prime }(\xi )\mathbf{v}_{s})(\mathbf{s}^{\prime
}(\xi )\frac{\delta E(\mathbf{s)}}{\delta \mathbf{s(}\xi ,t)})+\frac{\delta
E(\mathbf{s)}}{\delta \mathbf{s(}\xi ,t)}\mathbf{v}_{s}]\exp (-\frac{H(%
\mathbf{s)}}{k_{B}T})  \nonumber \\
\end{eqnarray}%
In parametrization where $\mathbf{s}^{\prime }(\xi )\delta E(\mathbf{s)/}%
\delta \mathbf{s(}\xi ,t)=0$ all terms vanish.

Thus, the reversible (not associated with dissipation) term in the original
equation of motion (\ref{ds/dt}) does not contribute to the functional
dynamics of the probability distribution. We say that it is the divergents
free term. It is understood, that the foregoing relates only to the case of
thermal equilibrium, i.e. valid only for solutions (\ref{Gibbs 1}).

~~~~ The last term in the integrand (\ref{FP Vns}) can be converted by using
the fluctuation-dissipation theorem (\ref{correlator}) as follows:
\begin{equation}
\int d\xi \int d\xi ^{\prime }\frac{k_{B}T\alpha }{\pi }~\delta (\xi _{1}%
\mathbf{-}\xi _{2})~\delta (t_{1}-t_{2})\delta _{\eta _{1\ ,}\eta _{2}}~%
\frac{\delta }{\delta \mathbf{s}(\xi ^{\prime })}\exp (-H\left\{ \mathbf{s}%
\right\} /k_{B}T)
\end{equation}%
Applying identities relation (\ref{B_trans}) and (\ref{dpv/ds}) we can
easily verify that the resulting expression exactly compensates the
dissipative term in the integrand in (\ref{FP Vns})(we recall that we work
in a frame where $\mathbf{v}_{n}$ is zero). This implies that the Gibbs
distribution with the Hamiltonian $H\left\{ \mathbf{s}\right\} $ (\ref%
{hamiltonian}) is indeed the solution of the Fokker--Planck equation, as it
must be in accordance with the general physical principles.

\section{Conclusion}

\label{Conclusion} Relations (\ref{Gibbs 1})-(\ref{Lamb(r)}) are assigned to
evaluate partition function and to calculate various statistical properties
of the vortex tangle. A considerable simplification in the evaluation of the
partition function can be reached with the use of the Edwards trick (see,
for details \cite{Edwards1979}, \cite{Kleinert1991}). Namely, the quantity $%
\exp (-\frac{E\{\mathbf{s\}-P(v_{n}-v_{s})}}{k_{B}T})$ can be written as a
Gaussian path integral over an auxiliary vector field $\mathbf{A(r})$, which
is readily evaluated. With the use of this technique we intend to calculate
the structure factors of quantum turbulence, e.g. average polarization of
the vortex loops composing the vortex tangle in the counterflowing helium
II, as well as anisotropy and mean curvature. These quantities were earlier
obtained only in the numerical work by Schwarz \cite{Schwarz1988}. The
results on properties of vortex tangle, which we plan to obtained on the
basis of the formalism, derived in the present work are supposed to compared
with numerous data on the quantum turbulence.

The work was supported by 15-02-05366  from RFBR (Russian Foundation of
Fundamental Research)

\end{document}